\pdfoutput=1
\documentclass[11pt,a4paper]{article}
\usepackage{amsmath,amssymb}
\usepackage{cite}
\usepackage{url}
\usepackage{mathtools}
\usepackage{mathbbol,amsmath}              % for math symbols.
\usepackage{graphics,graphicx,epsfig,ulem} 
\usepackage{subfigure}
\usepackage{feynmp}
\usepackage{slashed}
\renewcommand\[{\left[}
\renewcommand\]{\right]}

\def\sst{\scriptscriptstyle}

\def\bra{\langle}
\def\ket{\rangle}
\def\beq{\begin{equation}}
\def\eeq{\end{equation}}
\def\[{\begin{equation}}
\def\]{\end{equation}}
\newcommand{\C}[1]{\mathcal{#1}}

\newcommand{\startappendix}{
\setcounter{section}{0}
\renewcommand{\thesection}{\Alph{section}}
\renewcommand{\theequation}{\Alph{section}.\arabic{equation}}}

\newcommand{\Appendix}[1]{
\refstepcounter{section}
\begin{flushleft}
{\Large\bf Appendix: #1}
\end{flushleft}}

\begin{document}
\numberwithin{equation}{section}

\title{{\normalsize  \mbox{}\hfill IPPP/14/53, DCPT/14/106}\\
\vspace{2.5cm}
\Large{\textbf{Dark matter monopoles, vectors and photons}}}

\author{Valentin V. Khoze and Gunnar Ro\\[4ex]
  \small{ Institute for Particle Physics Phenomenology, Department of Physics} \\
  \small{ Durham University, Durham DH1 3LE, United Kingdom}\\[0.4ex]
    \small{  valya.khoze@durham.ac.uk, g.o.i.ro@durham.ac.uk}\\[0.8ex]
}

\date{}
\maketitle

\begin{abstract}
\noindent In a secluded dark sector which is coupled to the Standard Model via a Higgs portal interaction we arrange
for the existence of 't Hooft-Polyakov magnetic monopoles and study their implications for cosmology. We point out that 
a dark sector which can accommodate stable monopoles will also contain massless dark photons $\gamma'$ as well as
charged massive vector bosons $W'_{\pm}$. The dark matter in this scenario will be a combination
of magnetically and electrically charged species under the unbroken U(1) subgroup of the dark sector. We estimate the
cosmological production rate of monopoles and the rate of monopole-anti-monopole annihilation and conclude that monopoles
with masses of few hundred TeV or greater, can produce sizeable contributions to the observed dark matter relic density. 
We scan over the parameter space and compute the relic density for monopoles and vector bosons. Turning to dark
photons, we compute their contribution to the measured density of relativistic particles $N_{\rm eff}$ and also apply
observational constraints from the Bullet cluster and other large scale galaxies on long-range interactions 
of monopoles and of dark vector bosons. At scales relevant for dwarf galaxies we identify regions on the
parameter space where self-interacting monopole and vector dark mater components can aid solving the core-vs-cusp
and the too-big-to-fail problems.
\end{abstract}

%\centerline{\date{\today} }
\bigskip

\bigskip

$\quad$ JHEP 1410 (2014) 61
\thispagestyle{empty}
\setcounter{page}{0}

\newpage
\section{Introduction}
\label{sec:1}

A general and simple way to introduce dark matter into particle physics 
is to extend the Standard Model (SM) by a Dark sector, which includes dark matter particles, as well as possibly
many other degrees of freedom.

If the Dark sector contains magnetic monopoles, can they be dark matter?
The Dark sector, as the SM itself, can contain non-Abelian gauge fields and scalar fields, and if the Dark sector model is of the Georgi-Glashow
type, 't Hooft-Polyakov monopoles  \cite{'tHooft:1974qc,Polyakov:1974ek} will necessarily exist -- this by itself of course does not mean that
they have been or can be produced. These dark magnetic monopoles carry dark magnetic charge and
are stable; they can also be relatively light, for example much lighter than the monopoles of Grand Unified theories. 

The motivation of this paper is to investigate the cosmological consequences of magnetic monopoles of the dark sector.
Can the monopoles contribute to the observed relic density and how sizeable can their contribution be; what was the cosmological 
production rate of monopoles and their annihilation rate; more importantly, which additional features emerge from the dark sector for it to be able to 
support monopoles? What are the cosmological and phenomenological constraints on the complete model and is it more attractive than the more traditional 
models of collisionless WIMP dark matter?
These are some of the questions we want to address.

Models of dark matter with non-Abelian Dark sectors interacting with the visible SM sector only weakly, for example via some portal 
interactions is a popular approach to dark matter, see e.g. \cite{Strassler:2006im,Pospelov:2007mp,ArkaniHamed:2008qn}.
Cosmological production rate of light and heavy magnetic monopoles and their contributions to dark matter were discussed previously in
 \cite{Murayama:2009nj}. We will incorporate these results in our analysis.
More recently dark sector monopoles and vector bosons were considered in \cite{Baek:2013dwa} with the authors of \cite{Baek:2013dwa}
concluding that the monopole contribution to the dark matter density should be negligible.
This does not agree with our findings in Section 3.

\medskip

Multi-component vector and scalar dark matter resulting from a non-Abelian Dark sector coupled to the SM via the Higgs portal interaction,
was studied recently in \cite{Khoze:2014xha} also with the view of helping to stabilise the SM Higgs potential, 
following earlier work \cite{Hambye:2013dgv,Carone:2013wla}. We will now introduce magnetic monopoles into the theory.

\subsection{The model}
\label{sec:2}

Consider the Standard Model extended by a hidden (a.k.a. Dark) sector which contains an $SU(2)_{\sst D}$ gauge group and a scalar field $\Phi$ in the adjoint representation of 
$SU(2)_D$ (this is the simplest model of interest containing topologically stable monopoles; the Reader can imagine more complicated versions 
of the Dark sector(s), but the present model is fully adequate for our settings).
The Lagrangian for the Dark sector is:
\begin{equation}
\C{L}_{\sst D}= -\frac{1}{2} {\rm Tr}F'_{\mu\nu} F^{' \mu\nu} +
{\rm Tr}(D_\mu\Phi (D^\mu \Phi)^\dagger) -\lambda_\phi {\rm Tr}(\Phi \Phi^\dagger)^2 +m^2\, {\rm Tr}(\Phi \Phi^\dagger) \, , \quad \Phi = \phi_a \frac{\sigma_a}{2}.
\label{eq:LD}
\end{equation}
Here $F'_{\mu\nu}$ is the field strength of the $SU(2)_{\sst D}$ gauge field
$A'_\mu = A^{' a}_\mu \frac{\sigma_a}{2}$,
the covariant derivative is $D_\mu \Phi=\partial_\mu\Phi +i g_{\sst D}[A'_\mu,\Phi]$, where $g_{\sst D}$ is the gauge coupling, and $\sigma_{a=1,2,3}$ are the Pauli matrices. 

The $\Phi$-field also couples to the SM via the Higgs Portal interaction, 
\[
\C{L}_{\rm HP} = \lambda_{\rm P} (H^\dagger H) {\rm Tr}(\Phi \Phi^\dagger)\,.
\label{eq:HP}
\]
 In the absence of other matter fields in the Dark sector, these are the only interactions between 
the SM and the Dark sector. In particular, there is no kinetic mixing between the non-Abelian Dark sector $SU(2)_{\sst D}$ and the SM gauge groups.

The scalar potential in our Dark-sector Lagrangian \eqref{eq:LD} contains the negative mass-squared term, $-m^2 \, {\rm Tr}(\Phi \Phi^\dagger)$, for the adjoint scalar.
As the result, $\C{L}_{\sst D}$ has
a non-trivial vacuum $\left<\Phi\right> \neq 0$ which brakes the $SU(2)_{\sst D}$ gauge symmetry to $U(1)_{\sst D}$. Using gauge freedom we can set
\[
\left<\Phi\right>=\left<\phi_3\right>\frac{\sigma_3}{2}\,\, , \quad {\rm where} \quad \left<\phi_3\right>=w=m/\sqrt{\lambda_\phi}\,.
\]
After symmetry breaking in the Dark sector we get two massive gauge bosons $W_\pm'$ with mass $M_{W'}=g_{\sst D}w$, one massive scalar $m_\phi=\sqrt{2}m$ and one massless gauge boson $\gamma'$. $SU(2)_{\sst D}$ has been broken to a massless $U(1)_{\sst D}$.

\medskip
The effect of symmetry breaking is communicated from the Dark to the SM sector
via the Higgs Portal interaction \eqref{eq:HP} which can generate the $\mu^2_{\sst SM}$
 term in the SM effective potential,
\begin{equation}
V(H)_{\sst SM}=-\frac{1}{2}\mu^2_{\sst SM}H H^\dagger +\lambda_{\sst SM}(H H^\dagger)^2 \,.
\label{eq:Vsm}
\end{equation}
If  $\mu^2_{\sst SM}$ was absent at tree level, the Dark sector generates the contribution
$\mu^2_{\sst SM} = \lambda_{\rm P}\langle|\Phi|\rangle^2$
and triggers the electroweak symmetry breaking with the Higgs vev and mass,
\begin{equation}
v =\,  \frac{\mu_{\sst SM}}{(2\lambda_{\sst SM})^{1/2} }
 \,\,\simeq 246 \,{\rm GeV}\, , \qquad
m_{h\, \sst SM} = \mu_{\sst SM} \,\,\simeq 126\, {\rm GeV} \,.
\label{SMfirst}
\end{equation}

\subsection{Monopoles}
\label{sec:2.1}

It is well known that the spectrum of the SU(2) gauge theory with an adjoint scalar must contain 't Hooft-Polyakov magnetic monopoles
\cite{'tHooft:1974qc,Polyakov:1974ek}.

The mass of the monopoles is bounded from bellow by the Bogomolny bound \cite{Bogomolny:1975de},
\[
M_{\rm m}\, \geq\,   \frac{4\pi}{g_{\sst D}} \, w \, =\frac{M_{W'}}{\alpha_{\sst D}}\,,
\]
and monopoles have a magnetic charge $g_{\sst mD} =\frac{4\pi}{g_{\sst D}}$. For BPS monopoles in the limit where $\lambda_\phi \rightarrow 0$ the mass saturates the Bogomolny bound. More generally, away from the BPS limit, the monopole mass is given by
$M_{\rm m} = \frac{M_{W'}}{\alpha_{\sst D}}\,f(\lambda_\phi/g_{\sst D}^2)$ where $f$ is a smooth monotonically increasing function from 
$f(0)=1$ to $f(\infty) \simeq 1.787$, see e.g. \cite{Preskill:1984gd}. Consequently we will use the Bogomolny bound as a reasonable approximation for the monopole mass
for all values of $\lambda_\phi$.

\subsection{Mass-scale generation}
\label{sec:2.2}

What is the origin of the $m^2$ term in \eqref{eq:LD}? 

\medskip
(1.) We can choose to make the full theory classically scale-invariant (CSI), in this case all input mass scales of the classical
Lagrangian are set to zero, thus $m^2_{\rm cl} \equiv 0$.\footnote{We should make it clear that one can treat $m^2$
as an input parameter and not consider CSI at all, without invalidating any of the cosmological arguments that will follow.}
The vacuum expectation value $\left<\Phi\right> =w \neq 0$ is then
generated radiatively via the Coleman-Weinberg (CW) mechanism  \cite{CW}. In the Appendix 
we outline how this works in massless Georgi-Glashow theory.
The Dark gauge symmetry is broken by $\left<\Phi\right> $ and this can be recast as generating an effective $m^2$ term in \eqref{eq:LD} 
in the CSI Standard Model $\times \, SU(2)_{\sst D}$ theory.
This is a minimal scenario where dynamical mass generation occurs directly in the Dark sector, 
i.e. we have identified the mass-generating sector with the Dark sector, $SU(2)_{\sst CW} = SU(2)_{\sst D}$.

\medskip
(2.) A complimentary approach is to keep the mass-scale-generating sector and the Dark sector distinct. Then interactions between the 
two sectors would transmit the mass scale from the mass-scale-generating sector to the he Dark sector. For example,
in CSI settings we can think of the SM $\times \,SU(2)_{\sst D} \times G_{\sst CW}$ model, where $G_{\sst CW}$  is the Coleman-Weinberg gauge sector
which generates the vev $\langle \varphi_{\sst CW} \rangle$ for the CW scalar field. This radiatively generated scale is then transmitted to the Dark
sector scalar and to the SM Higgs field via scalar portal interactions, 
$\C{L}_{\rm Portal} \ni  \lambda_{\sst CWD}\,  |\varphi_{\sst CW}|^2 \,{\rm Tr}(\Phi \Phi^\dagger) 
+\lambda_{\sst CWH}\,  |\varphi_{\sst CW}|^2 \, (H^\dagger H) $ such that $  \lambda_{\sst CWD}\,  |\langle \varphi_{\sst CW}\rangle |^2 \,=m^2$ 
in \eqref{eq:LD}.\footnote{In this scenario the induced SM Higgs mass parameter in \eqref{eq:Vsm} is
 $\mu^2_{\sst SM} = \lambda_{\sst CWH}\,  \langle|\varphi_{\sst CW}|\rangle^2  + \lambda_{\rm P}\langle|\Phi|\rangle^2$.}
In the above, $G_{\sst CW}$ is an example of the mass-generating sector, in general it does not have to be reliant on the CW mechanism, the mass scale can 
arise from any dimensional transmutation-type dynamical argument, including a strongly coupled sector.

\medskip

Our reason for distinguishing between these two classes of models is the effect on the cosmological production of magnetic monopoles.
The monopole production rate \cite{Kibble:1976sj,Zurek:1985qw}
will depend on the nature of the phase transition in the Dark sector when the temperature in the early 
Universe falls below the critical temperature of $SU(2)_{\sst D}$.
In the Coleman-Weinberg sector the phase transition is of the first order, while in the Standard Model sector the electroweak phase transition 
is very weakly first order or second order~\cite{Anderson:1991zb,Dine:1992wr,Quiros:1999jp}.
The distinction can be traced to the value of the scalar self-coupling constant: in CW models $\lambda$ is small
relative to the gauge coupling (resulting in CW scalar masses being 1-loop suppressed relative to $W'$ masses), in the SM this is not the case, with the Higgs being heavier than $W$ and $Z$.

\bigskip
 
The Dark sector model \eqref{eq:LD} has three dark ingredients: dark photons $\gamma'$, dark massive vector bosons $W'_{ \pm}$,
and dark magnetic monopoles $M'_{{\rm mg} \pm}$. Massless $\gamma'$ photons is the Dark Radiation, it will be discussed below in section {\bf \ref{sec:3}}.
The remaining two ingredients, $W'_{ \pm}$ and $M'_{{\rm mg} \pm}$ are the two Dark Matter candidates in our model, they will be analysed in
section {\bf \ref{sec:4}}.
We will show  that the cosmological production of magnetic monopoles in the Dark sector is enhanced when the $SU(2)_{\sst D}$ phase transition is 
of the second order. Consequently, the monopoles  contribution to the observed Dark Matter relic density can be sizeable 
in models with a second order phase transition in the Dark sector. (For the models where the phase transition is strongly first order, it is unlikely.)
In section {\bf \ref{sec:5}} we will combine all three dark ingredients  and analyse the effect of the long-range forces acting on the 
Dark electromagnetic matter and compere with observations.
 
 We will show that both dark monopoles and dark vector bosons are viable dark matter components. The dark matter is not collisionless
 and hence we will check that it satisfies the known observational constraints. 
We will further argue that both monopole and vector components of dark matter give the transfer cross-sections 
of the right magnitude to be able to aid in solving the `too-big-to-fail' and the `core-vs-cusp' problems at dwarf galaxy scales. 

\bigskip

A viable framework for a fundamental particle theory beyond the Standard Model should address all the sub-Planckian shortcomings of the Standard Model.
This includes the generation of: baryon asymmetry, dark matter, and the electroweak scale (together with the stabilisation of the Higgs potential). The origin
of neutrino masses, and possibly the solution of strong CP problem and a particle mechanism of cosmological inflation are also on the list.
A  framework of BSM model building based on classical scale invariance has become popular in recent literature 
\cite{Bardeen:1995kv} 
as it can address many of these problems 
in predictive models with small numbers of parameters.
The high degree of predictivity/falsifiability arises from the fact that all mass scales have to be generated dynamically and, 
one cannot attempt to extend or repair the model by introducing new mass thresholds where new physics might enter.
The approach to self-interacting dark matter in the present paper is consistent with classical scale invariance (even though it is not required). 
In this case the origin of the dark matter scale (including the monopole mass and the vector boson mass) is tied to the SM electroweak scale
and to all other relevant scales of the full model.

\bigskip

Some additional applications of monopoles to dark matter physics were discussed in \cite{Sanchez:2011mf}
where TeV-scale monopoles in a hidden sector gave a decaying dark matter candidate due to a small kinetic mixing and a hidden photon mass. In our settings
there are no heavy messenger fields between the two sectors to induce the kinetic mixing and the monopoles are stable.
In Ref.~\cite{Fischler:2010nk}
it was pointed out that there is region of parameters in supersymmetric models where invisible monopoles can be the dark matter.
On the opposite side of the spectrum, Ref.~\cite{Evslin:2012fe} 
considered galaxy-sized Õt Hooft-Polyakov magnetic monopoles.

\medskip

\section{Dark Radiation and $\mathbf{N_{\rm eff}}$}
\label{sec:3}

The massless dark photon $\gamma'$ that remains after the breaking of $SU(2)_{\sst D}$ to $U(1)_{\sst D}$ is a new relativistic particle.
In this section we will determine the contribution of $\gamma'$  to
the effective number of relativistic degrees of freedom and apply experimental constraints.

\medskip
During both BBN and CMB the evolution of the Universe depends on the density of relativistic particles,\begin{equation}
\rho_{\rm rel}\,=\, g_\star(T)\, \times \frac{\pi^2}{30}\, T^4\,,
\end{equation}
and $g_\star$ counts the number of all relativistic degrees of freedom. Following standard notation, see e.g. Ref.~\cite{Feng:2008mu} for more detail, $g_\star$ is given by,
\begin{equation}
g_\star(T) \,=\sum_{m_i<T}C_i \, g_i \times \left(\frac{T_i}{T}\right)^4\,,
\label{eq:gstar}
\end{equation}
where the sum is over all degrees of freedom, $T_i$ and $m_i$ are the temperature and the mass of particle $i$, the coefficients are $C_i=1$ for bosons and $C_i=7/8$ for fermions, 
and $g_i$ denotes internal degrees of freedom (e.g. for SM photons $g_\gamma=2$ counting two transverse polarisations, and for each flavour of SM neutrino $g_\nu=2$). This expression is conventionally rewritten in terms of the effective number of neutrinos, $N_{\rm eff}$:
\begin{equation}
g_\star\, = g_\gamma + \frac{7}{8}\, g_\nu \, N_{\rm eff}\times \left(\frac{T_\nu}{T}\right)^4\,=\,2 \,+\, \frac{7}{8}\, 2 \,N_{\rm eff}\,\left(\frac{4}{11}\right)^{\frac{4}{3}}
\label{eq:gstarN}
\end{equation}
\begin{equation}
\Delta N_{\rm eff}\,\simeq\, 2.2\,\Delta g_\star
\label{eq:DNeff}
\end{equation}
In the Standard Model $N_{\rm eff}=3.046$ and any new relativistic particles would further add to it. 
Recently the Planck Collaboration found $N_{\rm eff}=3.30\pm 0.27$ at the time of recombination from a combination of CMB and Baryon Acoustic Oscillation data\cite{Ade:2013zuv}. The projected Planck limit is $\Delta N_{\rm eff}=0.044$.
There is also a limit of $N_{\rm eff}$ from Helium abundance at BBN (T=1MeV) $N_{\rm eff}= 3.24 \pm 1.2 $(95\%).

\medskip
The $\gamma'$ is a relativistic particle and contributes to $N_{\rm eff}$. 
If the dark photon was in thermal equilibrium with the SM photon, $T_{\gamma'}=T$, then Eq.~\eqref{eq:gstar} would give $\Delta g_\star=2$ leading to $ \Delta N_{\rm eff}\simeq 4.4$, 
which is ruled out by the Planck data. 

\medskip

However, this is not what happens in our case where the SM and the hidden Dark sector have no direct mediators and interact only via the Higgs portal; the
two sectors will
loose thermal contact after the $SU(2)_{\sst D}$ phase transition to the broken phase and before BBN. The interactions between 
dark photons and the SM will have to proceed through $\gamma'$ coupled to virtual $W'$ bosons which are coupled to virtual scalars $\phi$ which
have a small mixing with the SM Higgs through the Higgs portal coupling.
This interaction rate will be negligible with respect to the Hubble constant,
$\Gamma \, <\,  H = T^2/M_{\rm Pl}^\star$,
and the hidden sector will be colder than the SM. 

Following \cite{Feng:2008mu} we will model this situation in terms of
two sectors that have had the same temperature when all the degrees of freedom where relativistic, and then decoupled at temperature $T_D$. 
At the time of the measurement, the temperature is $T_M$ which is either at recombination or BBN.
Assuming that entropy is conserved within each sector we have,
\begin{equation}
\frac{g^h_{\star s}(T^h_{M})\, T^{h \,3}_{M}}{g^h_{\star s}(T_D)\, T_D^3}\, =\, \frac{g^{sm}_{\star s}(T_{M})\, T^{3}_{M}}{g^{sm}_{\star s}(T_D)\, T_D^3}\,,
\label{eq:entr}
\end{equation}
where the superscript $h$ refers to the hidden sector and $sm$ is the Standard Model. 
The number of relativistic degrees of freedom $g_{\star s}$ relevant for the entropy count is given by the expression ({\it cf.}~Eq.~\eqref{eq:gstar}),
\begin{equation}
g_{\star s} (T) \,=\sum_{m_i<T}C_i \, g_i \times \left(\frac{T_i}{T}\right)^3\,.
\label{eq:gSstar}
\end{equation}
In the hidden sector
$g^h_{\star s}$ counts only $\gamma'$ plus relativistic particles that will decay into $\gamma'$. Hence,
\[
g^h_{\star s}(T_D)=2+n \qquad  {\rm and} \qquad
g^h_{\star s}(T^h_{BBN})=g^h_{\star s}(T^h_{CMB})=2 \,,
\]
where $n$ denotes the number of relativistic particles in the hidden sector, in addition to 2 polarisations of $\gamma'$, 
at the time when the two sectors decouple (i.e. before the phase transition to the broken phase). 
The number of SM degrees of freedom at the decoupling temperature is
\[
g^{sm}_{\star s}(T_D)\, =\, 106.75\,,
\]
and at the time of measurements,
\begin{eqnarray}
g^{sm}_{\star s}(T_{BBN}) &=&
2_{\gamma}\,+\, \frac{7}{8}(4_{e^\pm} + (3\times 2)_{\nu})
\, =\, 10.75\,, \\
g^{sm}_{\star s}(T_{CMB})&=& 2_{\gamma}\,+\, \frac{7}{8} (3.046 \times 2)_\nu \times\frac{4}{11}\,=\,3.94\,.
\end{eqnarray}

From Eqs.~\eqref{eq:DNeff}, \eqref{eq:gstar}, \eqref{eq:entr}
we deduce $\Delta N_{\rm eff}$ at the time of measurement (BBN or CMB),
\begin{equation}
\Delta N_{\rm eff}(T_M)\,=\, 2.2\, \Delta g_\star (T_M)
\,=\, 2.2 \times g_{\gamma'} \times \left(\frac{T^{h}_M}{T_M}\right)^4
\, =\, 4.4 \times \left(\frac{g^h_{\star s}(T_D)}{g^h_{\star s}(T^h_{M})}\frac{g^{sm}_{\star s}(T_{M})}{g^{sm}_{\star s}(T_D)}\right)^{4/3}\,,
\end{equation}
so that,
\begin{equation}
\Delta N_{\rm eff}(T_{CMB})=0.022 \times (2+n)^{4/3}
\end{equation}
\begin{equation}
\Delta N_{\rm eff}(T_{BBN})=0.08 \times (2+n)^{4/3}
\end{equation}

In the model with only the dark photon in the hidden sector we would have $n=0$, leading to $\Delta N_{\rm eff}(T_{CMB})=0.05$ 
and $\Delta N_{\rm eff}(T_{BBN})=0.2$ 
very similarly to the result in \cite{Brust:2013ova}.

\medskip
We can now get a limit on the number $n$ of degrees of freedom in the Dark sector which annihilate into $\gamma'$.\footnote{Even if 
some of these particles have a relic density of the right order of magnitude to give the correct dark matter density, almost all of the entropy in the species 
will have been transferred since freeze out normally happens at $T=M/20$. The vector bosons $W_\pm'$ can decay to both $\gamma'$ and $\phi$. Since $\phi$ 
mixes with the SM Higgs, this entropy will leak to the Standard Model particles, which could effectively increase $g^{sm}_{\star s}(T_D)$. 
The fraction of the entropy transferred to $\gamma'$ is given by the branching ratio $\Gamma_{W_\pm'\rightarrow \gamma'}$ which is assumed 
to dominate over the entropy transfer to the SM.}
Since the neutral scalar $\phi$ does not
couple to the dark photon, the lowest value of $n$ we can have in the Georgi-Glashow Dark sector is $n=6$ given by the three polarisations of $W^{'\pm}$. 
Additional matter fields or higher rank gauge groups would increase $n$ appropriately.

From the Planck limit $\Delta N_{\rm eff} <0.8(2\sigma)$ at $T_{CMB}$ we get an upper limit
$n<14(2\sigma)$.
A stronger limit $n \lesssim 7$  follows from the data on Helium abundance at BBN.  We conclude that our $SU(2)_{\sst D}$ gauge theory with
an adjoint scalar is consistent with the current available constraints on $N_{\rm eff}$. At the same time additional degrees of freedom in the Dark sector are disfavoured.

\medskip
For the minimal case, $n=6$ arising from $W_\pm'$ contributions (and assuming that their entropy does not leak to the 
SM particles\footnote{Since $\phi$ mixes with the SM Higgs
there is some entropy exchange between the two sectors which can increase $g^{sm}_{\star s}(T_{CMB})$.}) 
our model predicts 
\begin{equation}
\Delta N_{\rm eff}(T_{CMB})=0.022 \times (2+6)^{4/3} = 0.35\,
\end{equation}
which could be ruled out by Planck measurements as the projected sensitivity in $\Delta N_{\rm eff}$ is $0.044$.

\medskip

\section{Dark Matter Relic Density}
\label{sec:4}

In our model there are two Dark Matter candidates. The massive gauge bosons $W'_{\pm}$ 
are carriers of the (dark) electric charge of the unbroken $U(1)_{\sst D}$ and as such  are stable. They provide a vector dark matter (VDM) candidate.
The dark magnetic (anti)-monopoles $M'_{{\rm mg}\pm}$  carry topological magnetic charge of $U(1)_{\sst D}$ and 
serve as a candidate for the monopole dark matter (MDM). 
The combined contribution of VDM and MDM should amount to (or not exceed) 
the observed total dark matter abundance $\Omega_{DM}h^2=0.1187\pm 0.0017$ measured by the Planck Satellite \cite{Ade:2013zuv}.

\subsection{Dark Gauge Bosons: Sommerfeld enhancement and relic density}
\label{sec:4.1}

\begin{figure}[t!]
\begin{center}
\vspace*{-0.4cm}
\includegraphics[width=0.9\textwidth]{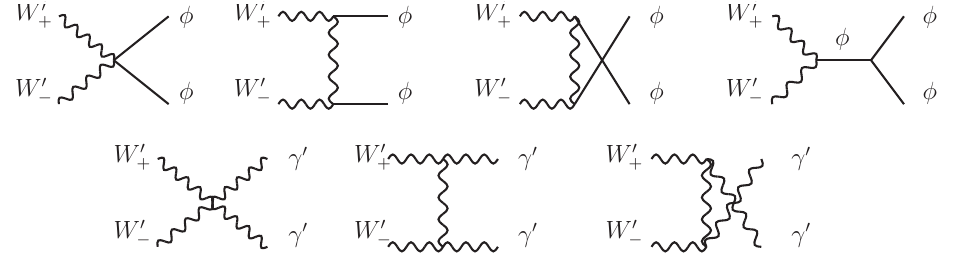}
\caption{Diagrams giving  dominant contribution to the $W'_{\pm}$ annihilation cross-section.}
\label{fig:feyn}
\end{center}
\end{figure}

$W'_{+}$ and $W'_{-}$ can annihilate into two dark photons $\gamma'$ or into two $\phi$ scalars. 
The dominant contribution to their annihilation is  given by the Feynman diagrams in Fig.~\ref{fig:feyn}.
Using these we have computed the leading order non-relativistic s-wave annihilation 
cross-section,\footnote{For simplicity, in the analytic expression on the {\it r.h.s.} of \eqref{eq:VDMphi}
we have assumed that $m_\phi \ll M_{W'}.$ We have checked that the inclusion of effects due to scalar masses does not 
make a noticeable change in our numerical results.
}
\begin{equation}
\left<\sigma v\right>_{\rm pert}\, =\, \frac{1579 \, g_{\sst D}^4}{2304 \pi  M_{W'}^2}\, -\,
\frac{5 \, g_{\sst D}^2 \, \lambda_\phi}{192 \pi  M_{W'}^2}\, +\, \frac{3 \, \lambda_\phi^2}{64 \pi  M_{W'}^2}\,.
\label{eq:VDMphi}
\end{equation}
This leading order perturbative cross-section is further enhanced at low velocities by the Sommerfeld effect 
\cite{Sommerfeld,Hisano:2002fk,Hisano:2003ec,Hisano:2004ds,Cirelli:2007xd},
which
arises from multiple dark photon exchanges in the $t$-channel between the incoming $W'_{+}$ and $W'_{-}$.
As the result we have,
\begin{equation}
\left<\sigma v\right>\, =\, S\, \left<\sigma v\right>_{\rm pert} \,,
\end{equation}
where the multiplicative Sommerfeld factor \cite{Sommerfeld,Cirelli:2007xd} is
\begin{equation}
S=\frac{\alpha_{\sst D} \pi}{v}\frac{1}{1-\exp\left[-\frac{\alpha_{\sst D} \pi}{v}\right]} \,,
\label{eq:S}
\end{equation}
and becomes relevant in the non-relativistic regime where the `perturbative' factor $\frac{\alpha_{\sst D} \pi}{v}$ is
no longer small. 

\begin{figure}[t!]
\begin{center}
%\vspace*{-1.cm}
\includegraphics[width=0.6\textwidth]{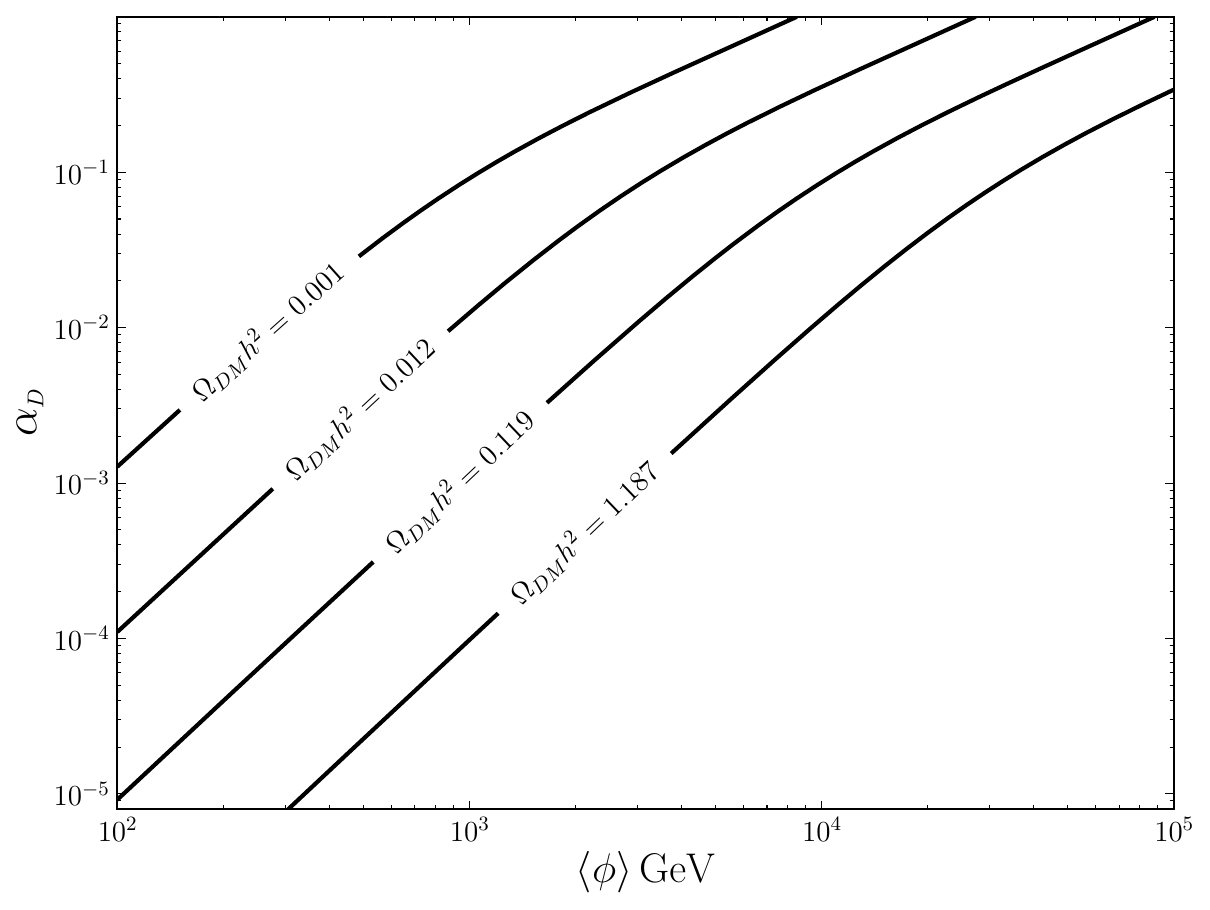}
\caption{Contours of the relic density of Vector Dark Matter.}
\label{fig:vdm}
\end{center}
\end{figure}

The relic density of vector dark matter is found by solving the Boltzmann equations,
\begin{equation}
\frac{d n_i}{dt}+3H n_i=-\left<\sigma v\right>(n_i^2-n_i^{eq \, 2})\,,
\end{equation}
where $n_i$ for $i=1,2$ is the density of $W'_+$ and $W'_-$ with $n_1=n_2$. Then the combined $W'_{\pm}$ density $n$ is twice that, $n=2n_1=2n_2$.
It satisfies the equation
\begin{equation}
\frac{d n}{dt}+3H n=-\left<\sigma v\right>_{\rm eff}(n^2-n^{eq\, 2})
\quad , \quad {\rm where} \quad 
\left<\sigma v\right>_{\rm eff} \,:=\frac{\left<\sigma v\right>}{2}
\,.
\end{equation}
Using this Boltzmann equation we can now write down the standard s-wave solution for the  dark matter abundance, see e.g. \cite{Ackerman:mha,Kolb:1990vq},
\begin{equation}
\Omega_{VDM}\, h^2 \, =1.07\times 10^9 \frac{x_f \, {\rm GeV}^{-1}}{(g_{\star s}/\sqrt{g_\star} )M_{\rm Pl}\, \left<\sigma v\right>_{\rm eff}}\, ,
\label{eq:VDMOm}
\end{equation}
where $x_f :=M_{W'}/T_f$ and $T_f$ is the freeze-out temperature. The expression for $x_f$ is
\begin{equation}
x_f\, =\, \log \left(0.038\frac{g}{\sqrt{g_\star}}M_{\rm Pl}M_{W'}\left<\sigma v\right>_{\rm eff}\right)
-\, \frac{1}{2}\log  \log \left(0.038\frac{g}{\sqrt{g_\star}}M_{\rm Pl}M_{W'}\left<\sigma v\right>_{\rm eff}\right)\,,
\label{eq:xf}
\end{equation}
where $g=6$ is the number of $W'_{\pm}$ degrees of freedom.

The relic density of $W'_{\pm}$  given by Eqs.~\eqref{eq:VDMOm}-\eqref{eq:xf}
is shown in Fig.~\ref{fig:vdm} on the two-dimensional plane $(\alpha_{\sst D}, w)$ of the Dark sector parameter space. 
In the CW case $\lambda_{\phi} \ll \alpha_{\sst D}$ and the scalar self-coupling $\lambda_{\phi}$ plays no role. We have also 
considered a more general case with $\lambda_{\phi} / \alpha_{\sst D}=$fixed, for example $=4$ similarly to the SM value, and continued to
scan over  $\alpha_{\sst D}$ and $\langle \phi\rangle$. We have found no noticeable difference in the relic density behaviour in Fig.~\ref{fig:vdm}.

The relic density curves in Fig.~\ref{fig:vdm} are seen to be bending at higher values of $\alpha_{\sst D}$. This is the consequence of the Sommerfeld enhancement factor in \eqref{eq:S}.
Indeed, while $S = 1$ in the perturbative regime $\frac{\alpha_{\sst D} \pi}{v} \ll 1$, its behaviour changes to $S = \frac{\alpha_{\sst D} \pi}{v}$
in the regime of larger gauge coupling or equivalently lower velocities -- which is precisely the reason for the bending. The velocity estimate is
$v = \sqrt{T_f/M_{W'}} = 1/\sqrt{x_f}$. In scanning over the parameter space in Fig.~\ref{fig:vdm} we found $x_f$ changing between 15 and 25 which gave 
the range of velocities $0.2 \lesssim v \lesssim 0.25$ in the Sommerfeld $S$ factor.

\medskip

\subsection{Dark Monopoles}
\label{sec:4.2}

\subsubsection{Production of Monopoles}
\label{sec:4.2.1}

Monopoles are topological defects which are produced during the phase transition in the early Universe. 
First we need to determine the order of the phase transition
of the $SU(2)_{\sst D}$ dark sector relevant for the monopole production. 
At sufficiently high temperature the only minimum of the effective potential of the Dark sector $V_{\sst D}(\phi,T)$ is at  the origin $\phi=0$ (here $\phi$ 
is the Dark sector scalar in the unitary gauge) and the $SU(2)_{\sst D}$ is restored.
As the Universe expands, the second minimum appears, and at critical temperature, $T=T_c$, the values of $V_{\sst D}$ in the two minima become equal. 
The phase transition is of the first order if 
there is a barrier separating the two minima at critical temperature. If, on the other hand, there is no potential barrier between the minima, the phase transition is of the second order. 

As already noted in section {\bf \ref{sec:2.2}} the character of the phase transition depends
on whether the Dark sector is of the CW-type or is distinct from it.
A simple estimate suffices to illustrate this point; to this end we proceed by writing the one-loop thermal potential (in the high-$T$ approximation) in the form
~\cite{Anderson:1991zb,Dine:1992wr}:
\[
V_{\sst D}(\phi,T)=D(T^2-T_0^2)\phi^2-E T \phi^3+\frac{\lambda_T}{4}\phi^4\,,
\]
with the parameters in our case (i.e. the model of \eqref{eq:LD}) given by,
\[D=\frac{g_{\sst D}^2}{4} , \quad
E=\frac{g_{\sst D}^3}{2\pi}, \quad 
T_0^2=\frac{1}{4D}\left(2m^2-\frac{3g_{\sst D}^4}{4\pi^2}w^2\right), \quad
\lambda_T=\lambda_{\phi}-\frac{3g_{\sst D}^4}{8\pi^2}\, \log \frac{g_{\sst D}^2w^2}{a_B T^2}\]
and $a_B \simeq e^{3.91}$. 
We get the critical temperature, $T_c$, when the values of $V_{\sst D}$ in the two minima become equal,
\[
T_c^2=\frac{T_0^2}{1-E^2/(\lambda_T D)}\ , \quad \phi_c=\frac{2ET_c}{\lambda_{T_c}}\,,
\]
with $\phi_c$ being the value of the field in the second minimum at this instance.
The strength of the first order phase transition is 
conventionally characterised by the dimensionless order parameter $\phi_c/T_c$,
which can be thought of as the separation between the two vacua in units of temperature.
We have
\[
\frac{\phi_c}{T_c}=\frac{2E}{\lambda_{T_c}}=\frac{g_{\sst D}^3}{\pi (\lambda_\phi+\frac{3g_{\sst D}^4 \times3.91}{8\pi^2})}\,.
\]
Strongly first order phase transitions
have $\phi_c/T_c \gtrsim 1$. The phase transition is weakly first order if the vacua at $0$ and $\phi_c$ are near
each other, and the phase transition changes from the very weakly first order to the second order for $\phi_c/T_c \ll 1$.

To have a second order phase transition we need:
\[
\frac{g_{\sst D}^3}{\pi} \ll \lambda_\phi+\frac{3g_{\sst D}^4 \times 3.91}{8\pi^2}
\]

At the same time, in the Coleman-Weinberg settings where all scales are generated radiatively so that the tree-level mass term is zero $m^2=0$, 
we have $\lambda_\phi=\frac{11g_{\sst D}^4}{8\pi^2}$ (as we recall in the Appendix). Hence the self-coupling $\lambda_\phi$ is small
%\[
%{\rm CW:}\qquad \frac{\phi_c}{T_c}\,=\,  \frac{1}{g_{\sst D}}\,\frac{8\pi}{ 22.73}\,=\, \frac{1}{ 3.2 \, \sqrt{\alpha_{\sst D}}} \gg 1 \quad {\rm for} \,\, \alpha_{\sst D} \ll 0.1 \,,
%\]
which should give the first order phase transition.
% gives the strongly first order phase transition for the weakly coupled CW sector, as expected.
This is indeed the case, but with the vanishing tree-level $m^2$ one can no longer use the high-temprerature approximation.
The recent analysis of the phase transition in this model carried out in \cite{Jaeckel:2016jlh} (section II) 
utilises the one-loop thermal effective potential for arbitrary values of $T$,
\begin{equation}
{\rm CW:}\qquad V_{T}(\phi)=\, \frac{3}{32 \pi ^2} \, g_{\sst \mathrm{D}}^4\, \phi^4
\left(-\frac{1}{2}+\log \left(\frac{\phi^2}{w^2}\right)\right) \, + \Delta V_{T}(\phi)\,,
\label{Vth0}
\end{equation}
where the $T$-dependent contribution is given by the well-known expression \cite{Dolan:1973qd} 
(and arises from $n=6$ degrees of freedom from the
three polarisations of the $W^{\prime \pm}$ bosons in the loop),
\begin{equation}
\Delta V_{T}(\phi)\,=\, 6\, \frac{T^{4}}{2\pi^{2}}\int_{0}^{\infty}\mbox{d}q\, 
q^{2}\log\left(1\mp\exp(-\sqrt{q^{2}+(g_{\sst \mathrm{D}}\phi)^{2}/T^{2}})\right)
\label{Vth1}
\end{equation}
It was shown in \cite{Jaeckel:2016jlh} that there is a barrier separating the two vacua
The value
of the critical temperature
where both minima are degenerate and the position of the second minimum were determined numerically to be at
\begin{equation}
{T_c} \simeq 0.4377\, {g_{\sst \mathrm{D}}\, \phi_c }\, , \quad {\phi_c} \simeq 0.85\, \phi_c\,,
\end{equation}
so that the order parameter  $\phi_c/T_c \simeq 1.94/g_{\sst \mathrm{D}} > 1$,  ensuring a first order phase transition.

\medskip

The production of magnetic monopoles in early Universe can be estimated using the Kibble limit \cite{Kibble:1976sj}. 
It is a lower limit on the density of magnetic monopoles created cosmologically, it is expressed in terms of the horizon volume, and it applies 
to both, 1st and 2nd order phase transitions,
\begin{equation}
\frac{n_{\rm m}}{T^3}\,  \geq  \, \left(\frac{T_c}{\sqrt{\frac{45}{4\pi^3g_\star}}M_{\rm Pl}}\right)^3
\label{eq:Kibble}
\end{equation}

First we justify this bound for the phase transition of the 2nd order \cite{Kibble:1976sj,Preskill:1984gd}.
During the phase transition the $\phi_a$ field changes from 0 to $|\phi|^2=w^2$. The direction of $\phi_a$ is the same inside a volume, 
$\zeta^3$, where $\zeta$ is the correlation length. At the critical temperature $\zeta$ diverges, but due to causality, information can only be exchanged inside the horizon.
The correlation length will be frozen in at the horizon scale $d_h \simeq H^{-1}$ and we will get a domain structure, with $\phi_a$ in different domains pointing in different directions. 
At domain intersection points the random orientation of the scalar field, given a non-trivial topology,
can give rise to magnetic monopoles with a probability $p$ close to 1.
We can estimate the density of monopoles created \cite{Preskill:1984gd}:
\begin{equation}
n_{\rm m}\, \propto \,p\,  \zeta^{-3} \sim\,  \zeta^{-3} \quad, \quad {\rm where} \quad
\zeta \, < \, d_h=H^{-1}\, ,
\end{equation}
and equation \eqref{eq:Kibble} follows.

If the phase transition in the Dark sector was of the first-order, a potential barrier is formed between the symmetric and the symmetry breaking vacua,
and below the critical temperature, the symmetric vacuum is meta-stable. Bubbles of the symmetry breaking vacuum will nucleate and expand. 
Inside each bubble the scalar field will have one random orientation. When the bubbles collide they can create magnetic monopoles. 
The density of magnetic monopoles will therefore proportional to the density of bubbles. Since the bubbles can not propagate faster than 
the speed of light the size of a bubble is limited by the horizon size. We therefore get a very similar bound \cite{Kolb:1990vq} on the density of magnetic 
monopoles as from the Kibble argument in Eq.~\eqref{eq:Kibble}, enhanced by a logarithmic factor \cite{Guth:1982pn}, 
\begin{equation}
{\rm 1st \ order \ ph. \,tr.:} \quad \frac{n_{\rm m}}{T^3}\,  \geq  \, \left[\frac{T_c}{\sqrt{\frac{45}{4\pi^3g_\star}}M_{\rm Pl}}
\, \log\left(  \frac{\sqrt{\frac{45}{4\pi^3g_\star}}M_{\rm Pl}} {T_c}\right)^4\,
     \right]^3 \,.
\label{eq:Kibble2}
\end{equation}

\medskip
Importantly, in the case of second-order phase transitions, the Kibble bound was refined by Zurek \cite{Zurek:1985qw} with an argument relying on a 
careful analysis of the timescales involved. 
The system undergoing the phase transition is characterised by the relaxation time $\tau$, and the correlation length $\zeta$,
\[
\tau\,=\,\frac{\tau_0}{\sqrt{|\epsilon(T)|}}  \, \,  , \quad {\rm and} \quad
\zeta\,=\, \zeta_0 \, |\epsilon(T)|^{-\nu}
\,,
\]
where 
\[
\epsilon(T):=\, \frac{T-T_c}{T_c}\,,
\]
and $1/2$ and $\nu$ are the critical exponents describing the degree of divergence of $\tau$ and $\zeta$  in the proximity 
of the critical temperature $T_c$. At the time $t$ close to the critical point $t_c$ one has $ t-t_c\, \propto \,  \epsilon(T) \,\to\, 0$ where the 
proportionality constant is the  quenching time-scale,
\[
\tau_Q :=\, \frac{t-t_c}{\epsilon(T)} \, \, .
\label{eq:tQdef}
\]
At the time $t_{\star}$ when the time interval to the critical point becomes equal to the relaxation time $\tau$, the system is no longer able to re-adjust. 
At this instance we have, 
\[
|t_{\star}-t_c| \,=\, \tau(t_{\star}) \,=\, \tau_0\, |\epsilon(t_{\star})|^{-1/2}\,,
\]
with the {\it l.h.s.} being via \eqref{eq:tQdef}  also  $=\tau_Q\, |\epsilon(t_{\star})|$, which implies that
\[
|\epsilon(t_{\star})|^{3/2} \,=\, \tau_0/\tau_Q\, \,  , \quad {\rm and} \quad
\zeta (t_{\star})\,=\, \zeta_0 \, |\epsilon(t_\star)|^{-\nu}\,=\, \zeta_0 \, |\tau_0/\tau_Q|^{2\nu/3}\,.
\label{eq:corrl}
\]
In our case $\tau_Q\,=\, H(T_c)^{-1}$ and for the remaining constants, from the Landau-Ginzburg theory one estimates \cite{Murayama:2009nj} that 
$\zeta_0 \simeq \tau_0 \sim 1/(\sqrt{\lambda_\phi} T_c).$ Classical value for the critical exponent $\nu$ is $1/2$ but quantum corrections can modify
this value.

The second equation in \eqref{eq:corrl} is the correlation length at the freeze-out temperature $t_\star$, it is the more accurate replacement of the Kibble-limit estimate
$\zeta \, < \, d_h=H(T_c)^{-1}$.

The monopole relic density from the Zurek mechanism today is then given by the following expression \cite{Zurek:1985qw,Murayama:2009nj}
(for conversion factors see Eqs.~\eqref{eq:conv1}-\eqref{eq:conv2}),
\[
{\rm 2nd \ order \ ph. \,tr.:} \quad
\frac{n_{\rm m}}{T^3}\, \simeq\, 
10^{-2}\, \left(\frac{M_{\rm m}}{\text{1 TeV}}\right)\left(\frac{30\,T_c}{M_{\rm Pl}}\right)^{\frac{3 \nu}{1+\nu}}\,,
\label{eq:ZM1}
\]
or (using numerical conversion Eqs.~\eqref{eq:conv1}-\eqref{eq:conv2}),
\begin{equation}
{\rm 2nd \ order \ ph. \,tr.:} \quad
\Omega_{\rm m}\, h^2\, = \, 1.5 \times 10^9\left(\frac{M_{\rm m}}{\text{1 TeV}}\right)\left(\frac{30\,T_c}{M_{\rm Pl}}\right)^{\frac{3 \nu}{1+\nu}}
\label{eq:ZM}
\end{equation}
Zurek's construction above is also valid for various condensed matter systems where the effect has been experimentally 
confirmed \cite{Ruutu:1998zz,Bowick:1992rz}.

The main difference between the Zurek result  \eqref{eq:ZM1}-\eqref{eq:ZM} and the Kibble lower limit \eqref{eq:Kibble} or
\eqref{eq:Kibble2}, is the power $p$ of the 
$(T_c/M_{\rm Pl})^p$ suppression
factor. It reduces from $p=3$ in the Kibble bound to the $p=3\nu/(1+\nu) \simeq p_{\rm cl} =1$ for $\nu_{\rm cl}=1/2$ in the Zurek bound. This makes it 
possible for relatively light monopoles with masses starting in a few hundred TeV range to contribute to dark matter, as can be inferred from
Fig.~\ref{fig:second_order_relic} in section {\bf \ref{sec:4.2.3}}. 
The Kibble bound would require 
monopoles to be at least in the $10^{11}$ GeV range or above to play a non-negligible role in the dark matter relic abundance, 
{\it cf.} Fig.~\ref{fig:first_order_relic}.

\subsubsection{Evolution of Monopoles}
\label{sec:4.2.2}

Magnetic monopoles are stable and can not decay due to conservation of their dark magnetic charge. The density of magnetic monopoles, once created, can therefore only be changed by monopole-anti-monopole annihilation.

In the diffusion approach \cite{Zeldovich:1978wj,Preskill:1979zi,Vilenkin},
the motion of monopoles in a plasma of electrically charged particles, in our case $W'_{\pm}$, is described by the Brownian walk with
thermal velocities $v_T = \sqrt{T/M_{\rm m}}$ and the mean free path $l_{\rm free}$, 
\[
l_{\rm free} \,=\, v_T \, t_{\rm free}\,=\, \sqrt{\frac{T}{M_{\rm m}}}\, \frac{M_{\rm m}}{T\sum_i n_i \sigma_i}\,,
\]
where  $\sigma_i$ is the classical cross-section for large angle scattering of a light particle with a monopole, 
\[
\sigma_i=\frac{g^2_{\sst mD}\, q^2_i}{(4\pi)^2 T^2}\,,
\]
$n_i$ is the number density and the sum is over all spin states.
The number density for relativistic particles is \cite{Kolb:1990vq}:
\begin{equation}
n_i=\,\frac{\zeta(3)}{\pi^2}\,T^3 \,,
\end{equation}
and for non relativistic particles of mass $M_i$ the number density is
\begin{equation}
n_i=\left(\frac{M_i\,T}{2\pi}\right)^{\frac{3}{2}} \exp\left(-\frac{M_i}{T}\right)\,.
\label{eq:nonrel}
\end{equation}
It is convenient to define the dimensionless quantity $B$,
\[
B\,:=\, T^{-1}\sum_i n_i \sigma_i \,\, , \quad {\rm so\ that:} \quad
l_{\rm free} \,=\, \frac{1}{B}\,\sqrt{\frac{T}{M_{\rm m}}}\, \frac{M_{\rm m}}{T^2}
\]
The attractive Coulomb force between the monopoles and anti-monopoles makes them drift towards each other 
during their random walk in the electric plasma. Their drift velocity is determined from the balance between the 
monopole-anti-monopole attraction and the drag force of the plasma. It is given by \cite{Vilenkin},
\[
v_{\rm drift}(r)\,=\, \frac{1}{B}\,\frac{g_{\sst mD}^2}{T^2 \,r^2}\,.
\]
Monopoles drift toward anti-monopoles through the plasma, the drag force dissipates monopole energy,
 and if the mean free path is less than the capture radius,
\[
l_{\rm free}\, \le \, l_{\rm capt} = g_{\sst mD}^2/(4\pi T)\,,
\]
the monopole-anti-monopole bound state is formed which ultimately annihilates to the ordinary elementary states.
The relevant time scale for the formation of the bound state is $t_{\rm drift} = r/v_{\rm drift}= 1/\Gamma_{\rm drift}$.
Therefore, the monopole-anti-monopole annihilation cross-section is given by,
\[
\sigma \,=\, \frac{\Gamma_{\rm drift}}{n_{\rm m}} \,=\, \frac{v_{\rm drift}(r)}{n_{\rm m} r}\,=\, 
\frac{1}{B}\,\frac{g_{\sst mD}^2}{T^2 }\,.
\label{eq:sigmmba}
\]

The resulting density of monopoles after annihilation is determined by the Boltzmann equation \cite{Preskill:1979zi},
\begin{equation}
\frac{d}{dx} \frac{n_{\rm m}}{s}=\frac{\sigma }{H(x)\,x} \left(\frac{n_{\rm m}}{s}\right)^2\,,
\quad {\rm where} \quad x:=\frac{M_{\rm m}}{T}
\end{equation}
with $\sigma$ on the right hand side given by \eqref{eq:sigmmba}.
The solution is known analytically \cite{Preskill:1979zi},  it quickly becomes independent of the initial conditions at $x_0$, 
resulting in,
\begin{equation}
\frac{n_{\rm m}}{s} (x)
\, \simeq\, \frac{2\pi B}{g_{\sst mD}^2}\,\frac{\sqrt{\frac{45}{4\pi^3g_\star}}M_{\rm m}}{M_{\rm Pl}}\, \,\frac{1}{x}
\end{equation}

If, following \cite{Preskill:1979zi}, we assume that the the plasma consists of particles that are relativistic from $x_0$ to $x_f$ where $x_f$ 
corresponds to the temperature where $l_{\rm free}$ and  $l_{\rm capt}$ become equal,
\begin{equation}
x^{-1}_f=\left(\frac{4\pi}{g_{\sst mD}^2}\right)^2\frac{1}{B^2}\,
\end{equation}
the result for the final number density of monopoles is in agreement with \cite{Preskill:1979zi}, 
\begin{equation}
\frac{n_{\rm m} }{s} (x_f)
\, \simeq\,
 \frac{2\pi}{Bg_{\sst mD}^2}\left(\frac{4\pi}{g_{\sst mD}}\right)^2\frac{\sqrt{\frac{45}{4\pi^3g_\star}}M_{\rm m}}{M_{\rm Pl}}\,.
\end{equation}
This diffusive capture process is effective only as long as the mean free path is smaller than the capture radius. 
At lower temperatures, where $l_{\rm free}$ exceeds  $l_{\rm capt}$, the rate of monopoles-anti-monopole annihilation cannot
compete with the Universe expansion and the monopole density freezes out at the value at $x_f$.

\medskip
There is an important difference between the more standard application of the diffusion method described above, where GUT monopoles 
were propagating in the plasma of very light relativistic electrons and positrons, and our model.
In our case the plasma is made up of $W_\pm'$ with masses $M_{W'}=g_{\sst D}\left<\phi\right>$ much closer to the monopoles
of the same dark sector.
Thus, the particles in the plasma will become non-relativistic fairly soon after the phase transition, when 
\begin{equation}
x_{\rm nr}=\frac{M_{\rm m}}{T_{\rm nr}}=\frac{M_{\rm m}}{M_{W'}}=\frac{1}{\alpha_{\sst D}}
=\frac{4\pi}{g_{\sst mD}^2}\,.
\end{equation}
After $x_{\rm nr}$ the density of the plasma will decrease exponentially, as per \eqref{eq:nonrel},
and the mean free path will therefore exponentially increase. The final monopole density in our model will thus be cut-off at $x_{\rm nr}$
\begin{equation}
\frac{n_{\rm m} }{s}(x_{\rm nr})
\, \simeq\,
\frac{B}{2}\,\frac{\sqrt{\frac{45}{4\pi^3g_\star}}M_m}{M_{pl}}\,.
\label{eq:mmbarfin}
\end{equation}

\subsubsection{Current density of Monopoles}
\label{sec:4.2.3}

\begin{figure}[t!]
\begin{center}
\vspace*{-1.cm}
\includegraphics[width=0.7\textwidth]{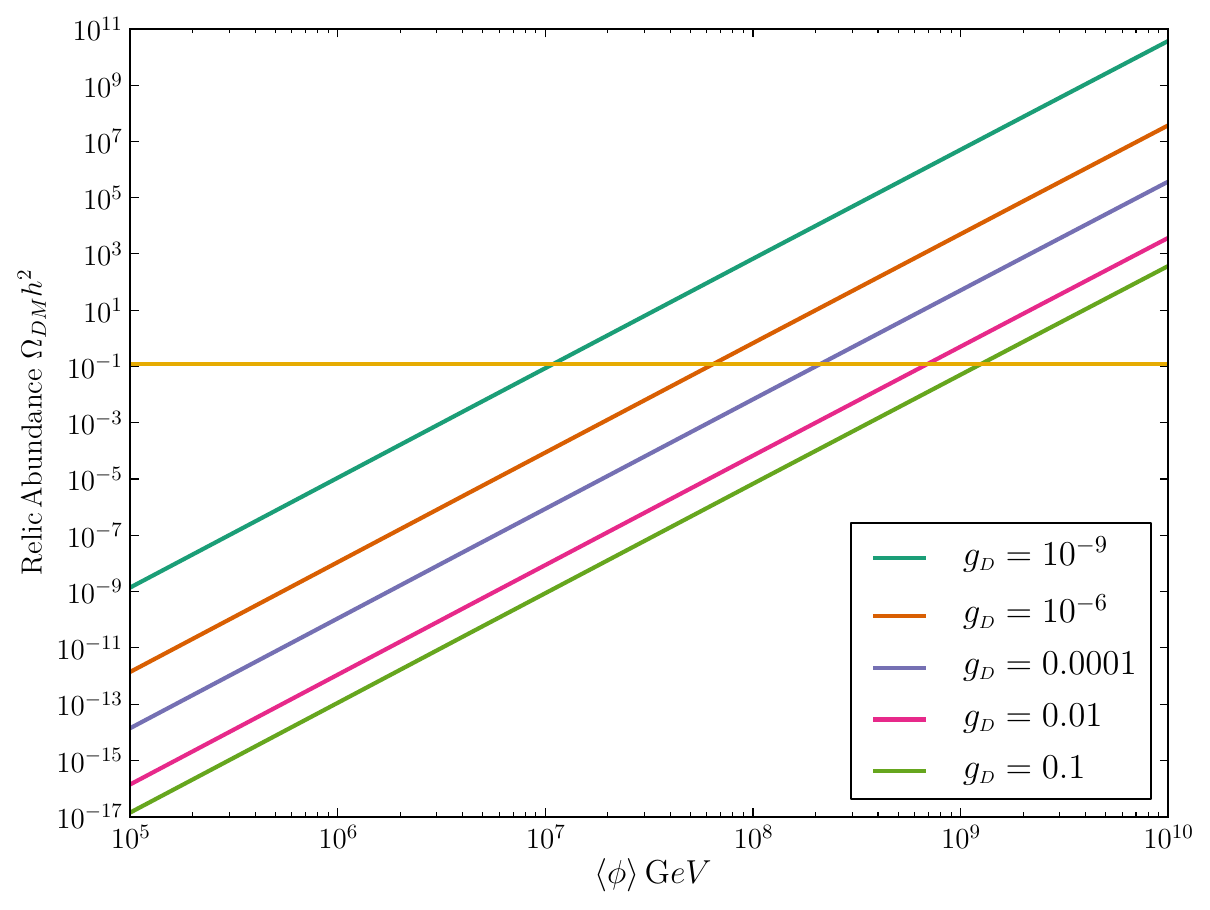}
\caption{The relic density of monopoles produced cosmologically during a first order phase transition as a function of the dark scalar vev
 $w=\langle \phi\rangle$ and for different values of the dark gauge coupling  $g_{\sst D}$.}
\label{fig:first_order_relic}
\end{center}
\end{figure}

To determine the current density in monopoles we first have to determine the type of the Dark sector phase transition and 
compute the initial monopole production density accordingly. If the initial production density is lower than the estimated density 
after monopole-anti-monopole annihilation \eqref{eq:mmbarfin},
the effect of annihilations is unimportant and the initial monopole density survives. If on the other hand the initial density is higher than the annihilation
density, the final monopole density is set by monopole-anti-monopole annihilations expression.

The conversion from monopole density, $n_{\rm m}/s$ or $n_{\rm m}/T$, to $\Omega_{\rm m}h^2$ is standard, 
\begin{eqnarray}
\Omega_{\rm m}h^2 &=&\rho_{\rm m} \,\, \frac{1}{\rho_{\rm crit}h^{-2}}\,,\\
\nonumber\\
\rho_{\rm m}h &=&\frac{n_{\rm m} }{s} \, M_{\rm m} \, s_0\, =\, \frac{n_{\rm m}}{T^3}\, M_{\rm m}\,T_0^3\,,
\end{eqnarray}
where subscript $0$ refers to the current time or temperature and the normalisation factors are given by,
\begin{eqnarray}
 \rho_{\rm crit}h^{-2} &=&1.9\times 10^{-29}{\rm g cm}^{-3}=7.53\times 10^{-47} \, {\rm GeV}^{4}\,,\\
s_0&=&\frac{2\pi^2}{45}g_\star(t=t_0)\, T_0^3\,,
\end{eqnarray}
with $T_0 = T_{CMB} = 2.73 \,{\rm K} = 2.35 \times 10^{-13} \, {\rm GeV}$ and $g_\star(t=t_0) =2$ in the Dark sector and 3.94 in the SM.
Thus
\begin{eqnarray}
\Omega_{\rm m}h^2 &=& \frac{n_{\rm m} }{s} \, \times \frac{M_{\rm m}}{1 \, {\rm TeV}} \times 1.5 \times 10^{11}\,,
\label{eq:conv1}\\
&=& \frac{n_{\rm m} }{T^3} \, \times \frac{M_{\rm m}}{1 \, {\rm TeV}} \times 1.7 \times 10^{11}\,.
\label{eq:conv2}
\end{eqnarray}

\begin{figure}[t!]
\begin{center}
\vspace*{-0.5cm}
\includegraphics[width=0.7\textwidth]{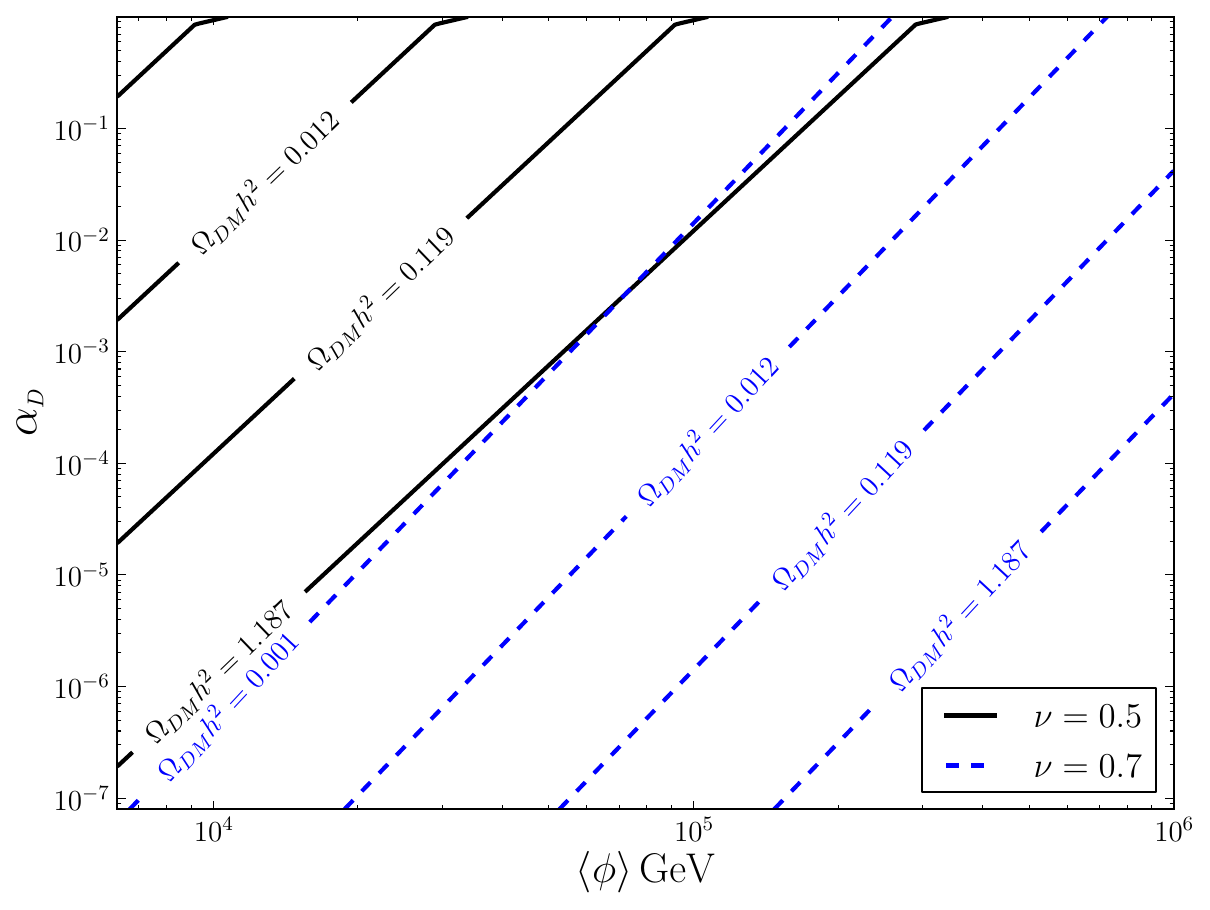}
\caption{The relic density of monopoles after a second order phase transition. Results are shown on the Dark sector gauge-coupling--vev plane 
for two different values for the critical exponent, $\nu=0.5$ (in black) and $\nu=0.7$ (in blue).}
\label{fig:second_order_relic}
\end{center}
\end{figure}

 The current relic density of monopoles for a first order phase transition
 computed using \eqref{eq:Kibble2}, is shown in Fig \ref{fig:first_order_relic}. We see that relic density depends strongly on
 the dark scalar field vev $w=\left<\phi\right>$ as this sets both the mass of the monopoles and the critical temperature of the phase transition. The density increases with lower coupling $g_{\sst D}$ as the mass of the monopoles increase.

\medskip

The current relic density for a second order phase transition, based on
\eqref{eq:ZM1}-\eqref{eq:ZM} combined with \eqref{eq:mmbarfin},
is plotted in Fig.~\ref{fig:second_order_relic} for two values of the critical exponent, $\nu=0.5$ and $\nu=0.7$.

%%%%%%%
%%%%%%%

\begin{figure}[t!]
\begin{center}
\vspace*{-1.cm}
\includegraphics[width=0.7\textwidth]{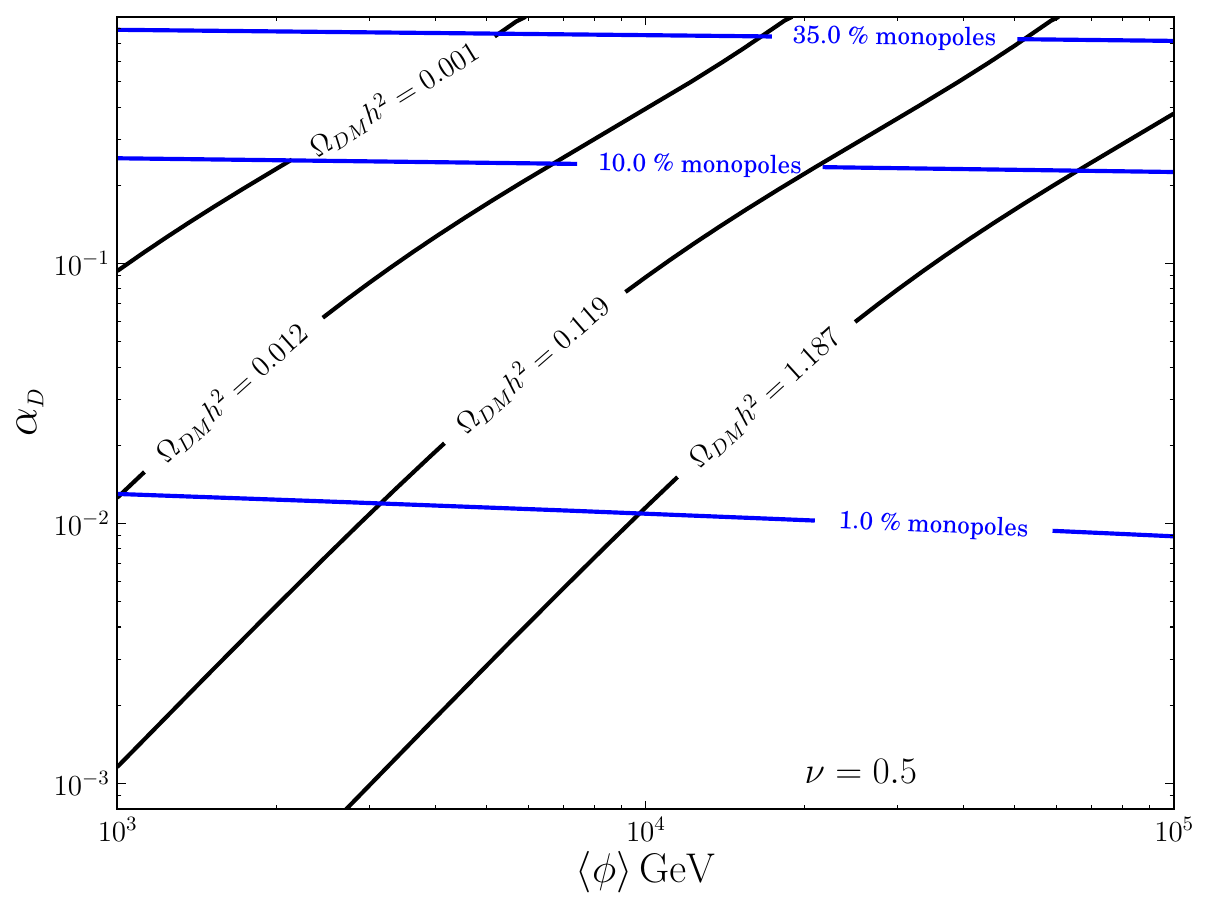}
\caption{Combined relic density of vector and monopole components of dark matter after a second order phase transition with
 the critical exponent $\nu=0.5$. The blue lines show the relative fraction of monopoles.}
\label{fig:combined_05}
\end{center}
\end{figure}
\begin{figure}[t!]
\begin{center}
\vspace*{-0.4cm}
\includegraphics[width=0.7\textwidth]{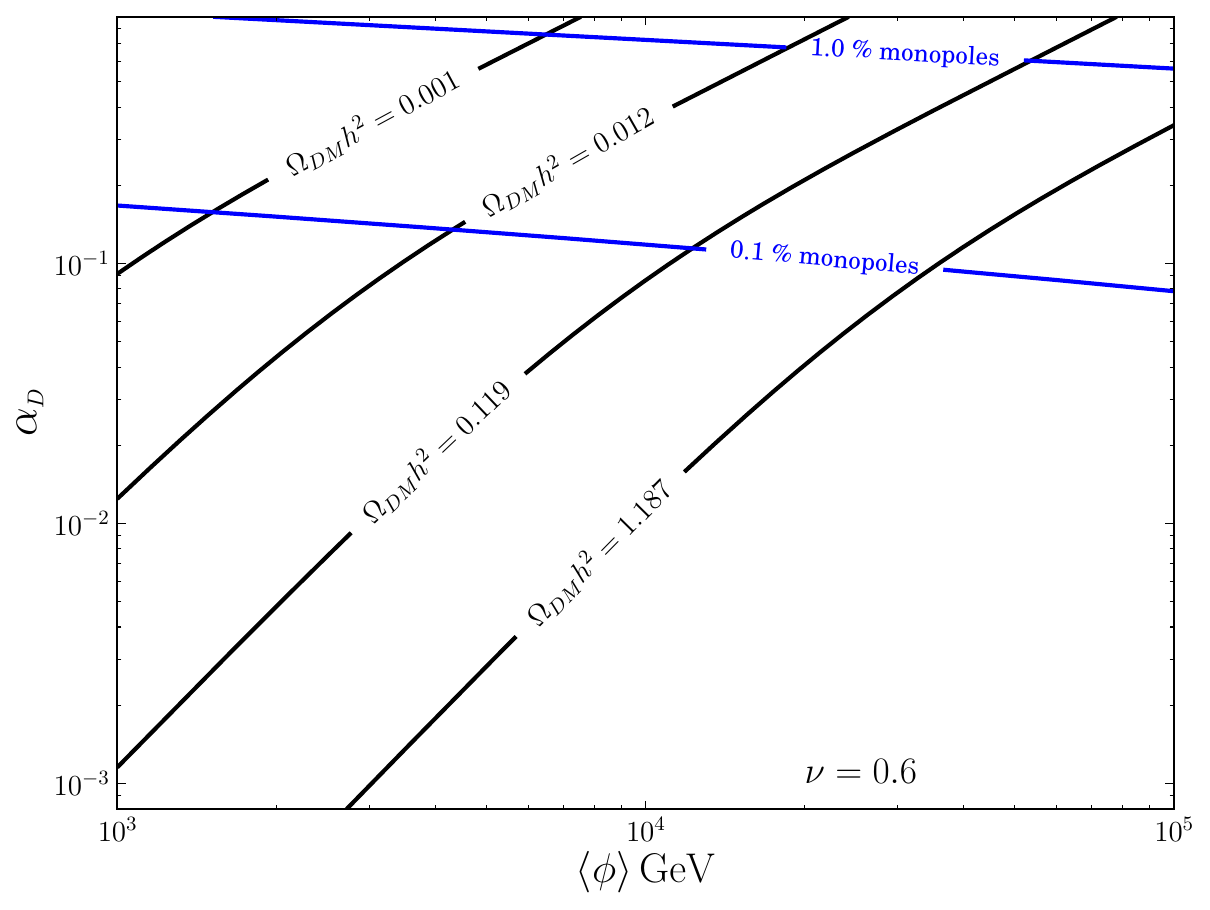}
\caption{Combined relic density for monopole and vector dark matter with $\nu=0.6$. Blue contours show the fraction of monopoles alone.}
\label{fig:combined_06}
\end{center}
\end{figure}

For a second order phase transition we can see that we have two components of dark matter both with a significant fraction of the observed relic density. The combined relic density can be seen in Figure \ref{fig:combined_05} for $\nu=0.5$ and in Figure~\ref{fig:combined_06} for $\nu=0.6$.

{\it Note Added:} Dark sector monopoles and vector bosons were also considered recently in Ref.~\cite{Baek:2013dwa} 
which concluded that the monopole contribution to the dark matter density is several orders of magnitude smaller than the observed dark matter relic density, and thus should be negligible. 
This  does not agree with our findings in Figs.~\ref{fig:combined_05}-\ref{fig:combined_06}.
 In computing the relic abundance of magnetic monopoles in the dark sector, the authors of
\cite{Baek:2013dwa} -- ArXiv version 1 -- appear to have concentrated only on a narrow patch of the available parameter space where 
both the electric $g_{\sst D}$  and magnetic $g_{\sst mD} =\frac{4\pi}{g_{\sst D}}$  coupling constants of the dark sector are not strongly coupled, $g_{\sst mD}  \sim g_{\sst D}  \sim 1$.  Our analysis, on the other hand, only requires a weakly-coupled electric theory formulation. The 't Hooft-Polyakov monopoles in this regime continue to be well described by the semi-classical theory in this  weak electric coupling regime. In the ArXiv version 2 of Ref.~\cite{Baek:2013dwa} which has appeared after our paper, the earlier restriction of being in
the weak magnetic coupling regime was lifted. However the authors of Ref.~\cite{Baek:2013dwa}  have still concluded based on their Fig. 8
that to achieve the monopole abundance of  10\% of the observed dark matter relic density would require what they describe as a 
$10^{-2}\, \%$ fine-tuning. Our results in Figs.~\ref{fig:combined_05}-\ref{fig:combined_06} do not  support such a conclusion. 
It follows from our Fig.~\ref{fig:combined_05} that the contour of the observed relic density
value $\Omega_{\rm DM} h^2 = 0.119$ can readily intersect the $10 \%$ monopole abundance contour,and even the $35 \%$ 
monopole abundance contour, and so on.

\section{Self-Interacting Dark Matter}
\label{sec:5}

Due to the unbroken $U(1)_{\sst D}$ symmetry we will have long-raged forces acting on the dark matter particles. Vector dark matter is electrically charged 
under the $U(1)_{\sst D}$ while the magnetic monopoles have magnetic charges. This self-interacting dark matter provides a framework which can solve 
cosmological problems of collisionless cold dark matter (CCDM) at small scales \cite{Spergel:1999mh}.  Numerical simulations \cite{Navarro:1996gj} 
based on CCDM
are very successful in describing the large scale structure of the Universe at scales $\gg 1$ Mpc.
However observations on galactic and subgalactic scales $\lesssim$ 1 Mpc are in conflict with the structure formation predicted by such simulations
\cite{Navarro:1996gj,Moore:1999gc}.

Collisionless dark matter predicts that density distributions of dwarf galaxy halos should have  a cusp in the centre, 
while observationally flat cores have been found; this is the core-vs-cusp problem. Cold dark matter simulations also predict too many 
too large sub-halos in the Milky Way. In particular, simulations which use collisionless dark matter predict ${\cal O}(10)$ 
sub-halos with velocities $v> 30$ km/s, but no halos have been observed with $v>25$ km/s. 
This is known as the `too-big-to-fail' problem. 

In order to address these problems with small scale structure models of 
self-interacting dark matter have been proposed and studied in recent literature.
References.~\cite{Buckley:2009in,Loeb:2010gj,Vogelsberger:2012ku,Zavala:2012us,Tulin:2013teo,Buckley:2014hja}
considered long-range Yukawa interactions between cold dark matter mediated by a light vector or scalar boson. 
The effects of an unbroken $U(1)$ symmetry with a massless force carrier were considered in
\cite{Ackerman:mha,Feng:2009mn,Kaplan:2009de}.

The result of self-interactions is to transfer energy between the dark matter particles. This effect is normally captured by the transfer cross-section
defined by,
\begin{equation}
\sigma_{\rm T}\,=\, \int d\Omega \, (1-\cos\theta)\, \frac{d\sigma}{d\Omega}\,,
\end{equation}
where $d\sigma/d\Omega$ is the usual differential cross-section. Even though our model contains a microscopically massless force carrier $\gamma'$,
in a plasma it is described by the Yukawa potential,
\begin{equation}
V(r)=\frac{\alpha_e}{r}e^{-m_{\gamma'} r}\,,
\end{equation}
where the  effective mass of $\gamma'$ is due to its interactions with the plasma and is given by the inverse of the Debye length $l_{\rm D}$,
\begin{equation}
m_{\gamma'}\,=\,\frac{1}{l_{\rm D}}\,=\,\frac{\left(4\pi \alpha_{\sst D} \rho\right)^{1/2}}{M_{\rm DM}\, v}\,.
\end{equation}
Here $\rho$ is the dark matter density in a galaxy and $v$ is its velocity. Since the density $\rho$ is small, the effective mass 
$m_{\gamma'}$ will be small and we can use the classical Coulomb limit $M_{\rm DM}v/m_{\gamma'} \gg 1$
 for both the attractive and repulsive potential 
with the result \cite{PhysRevLett.90.225002,PhysRevE.70.056405,Tulin:2013teo},
\begin{equation}
\sigma_{\rm T}=\frac{16\pi \alpha_{\sst D}^2}{M_{\rm DM}^2\,v^4}\, \log \left(1+\frac{M_{\rm DM}^2\, v^2}{2\alpha_{\sst D} m_{\gamma'}^2}\right)\,.
\end{equation}

If the energy transfer is large enough, self interacting dark matter could flatten out the cores of dwarf galaxies and decrease the number of large subhaloes, solving the core-vs-cusp and the too-big-too-fail problems. On the other hand if the cross section is too large the effects could be seen on larger scales and would be ruled out.

The limits on this cross-section come from comparing observations to simulations. One obvious constraint is from the Bullet cluster 
which gives an upper limit on the cross-section,
$\sigma_{\rm T}/M_{\rm DM}<1.25$ cm$^2$/g \cite{Randall:2007ph}. Since the transfer cross-section is very strongly velocity-dependent, 
it is important that this 
bound is imposed in the relevant velocity range $v\sim 1000$ km/s. 
There are also constraints of $\sigma_{\rm T}/M_{\rm DM}\lesssim$ 0.1 to 1 cm$^2$/g
from Milky Way scales in the velocity range of 200 km/s \cite{Tulin:2013teo}.

\medskip

\begin{figure}[t!]
\begin{center}
\vspace*{-1.4cm}
\includegraphics[width=0.7\textwidth]{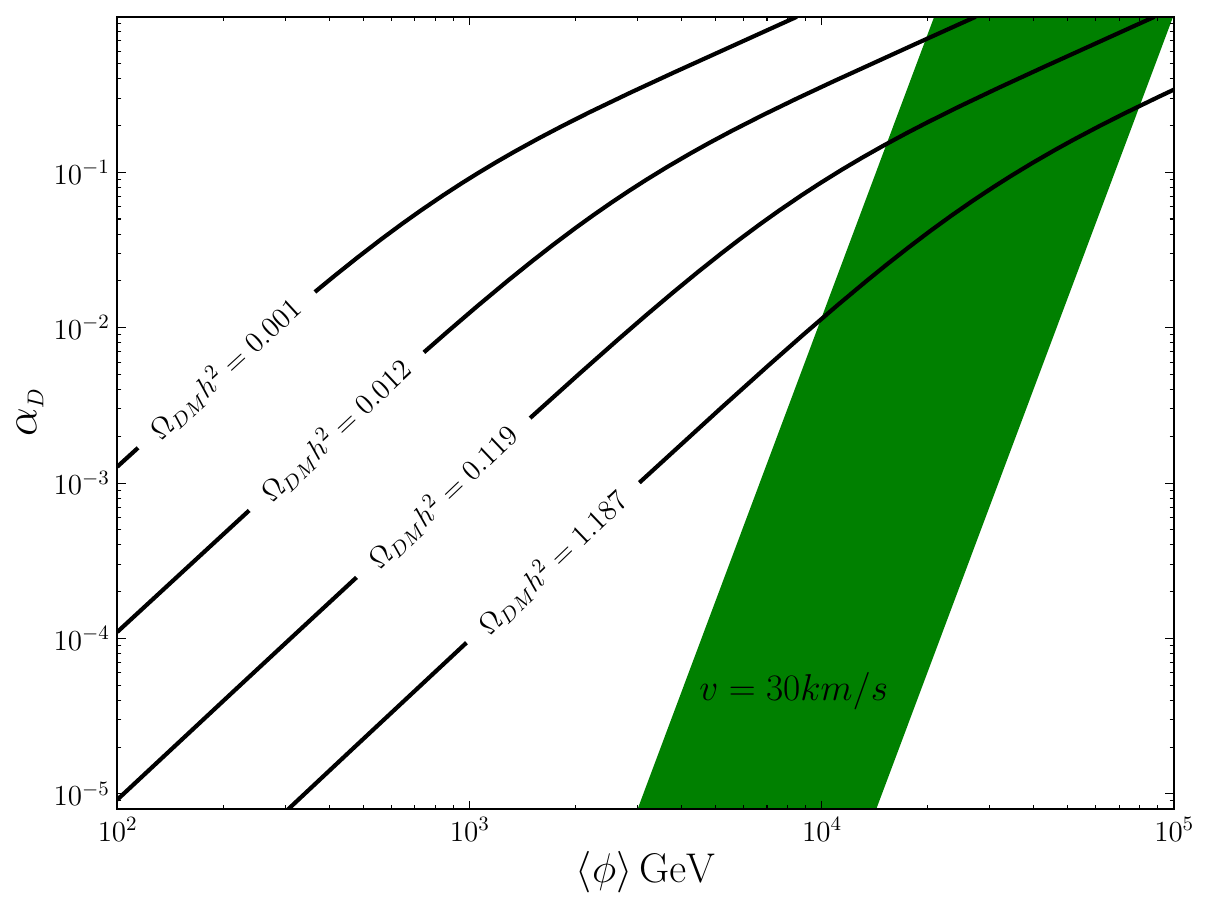}
\caption{Vector Dark Matter transfer cross-section and the relic density. The green region shows the region in parameter space
where $\sigma_T/m_{DM}$ is in the interval between 0.1 and 10 cm$^2$/g at velocity $v=30$ km/s relevant for solving the
core-cusp problem and the too-big-too-fail problem.}
\label{fig:vdm_limit}
\end{center}
\end{figure}

\begin{figure}[t!]
\begin{center}
\vspace*{-0.4cm}
\includegraphics[width=0.7\textwidth]{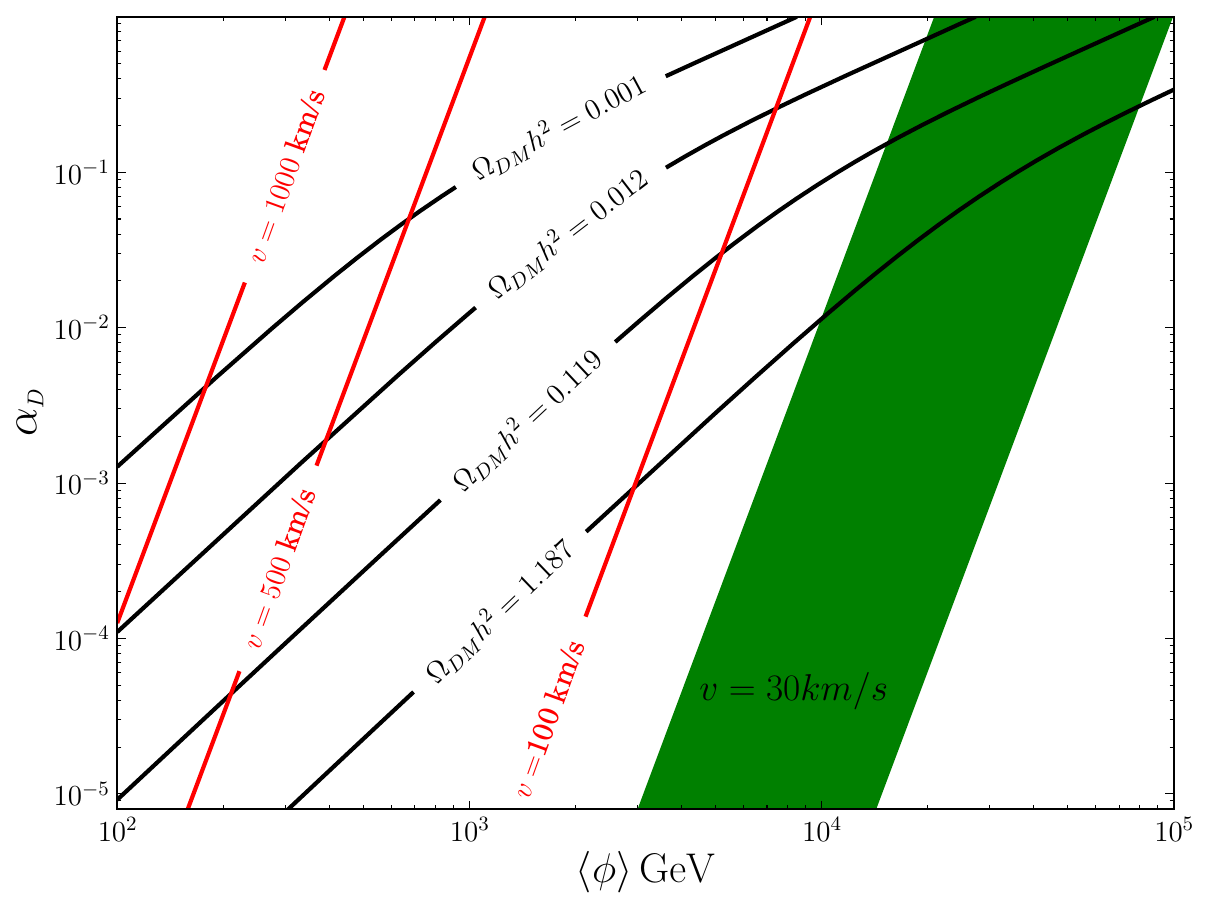}
\caption{Same as in Figure \ref{fig:vdm_limit} but with additional contours (in red) showing $\sigma_T/m_{DM}=1$ at higher velocities: $v=100$km/s, $v=500$km/s and $v=1000$km/s.}
\label{fig:vdm_limit2}
\end{center}
\end{figure}

\begin{figure}[t!]
\begin{center}
\vspace*{-1.4cm}
\includegraphics[width=0.7\textwidth]{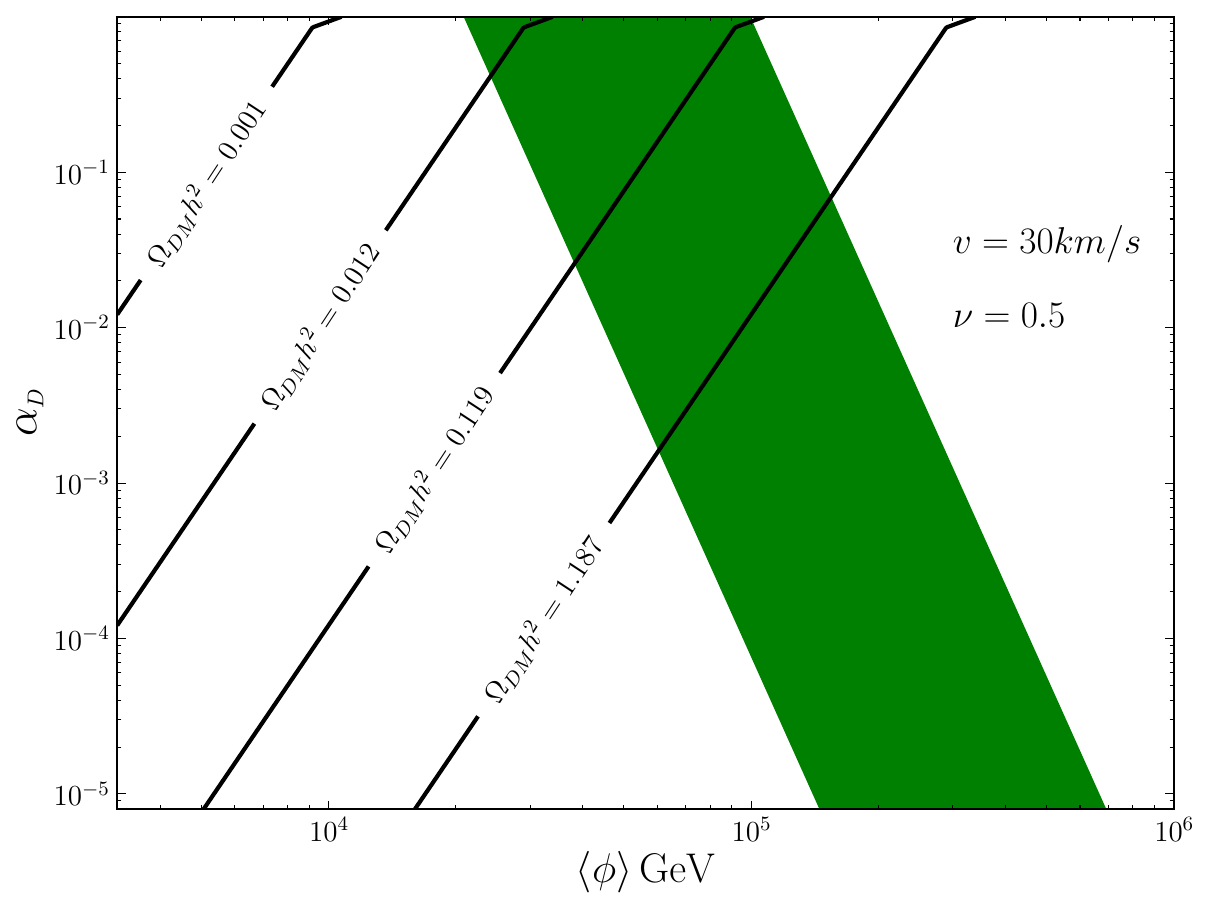}
\caption{Monopole Dark Matter transfer cross-section and the relic density contours for the critical exponent $\nu=0.5$. 
The region in green shows the region in parameter space
where $\sigma_T/m_{DM}$ is in the interval between 0.1 and 10 cm$^2$/g at velocity $v=30$ km/s relevant for solving the
core-vs-cusp problem and the too-big-too-fail problem.}
\label{fig:monopole_limit_05}
\end{center}
\end{figure}
\begin{figure}[t!]
\begin{center}
%\vspace*{-0.4cm}
\includegraphics[width=0.7\textwidth]{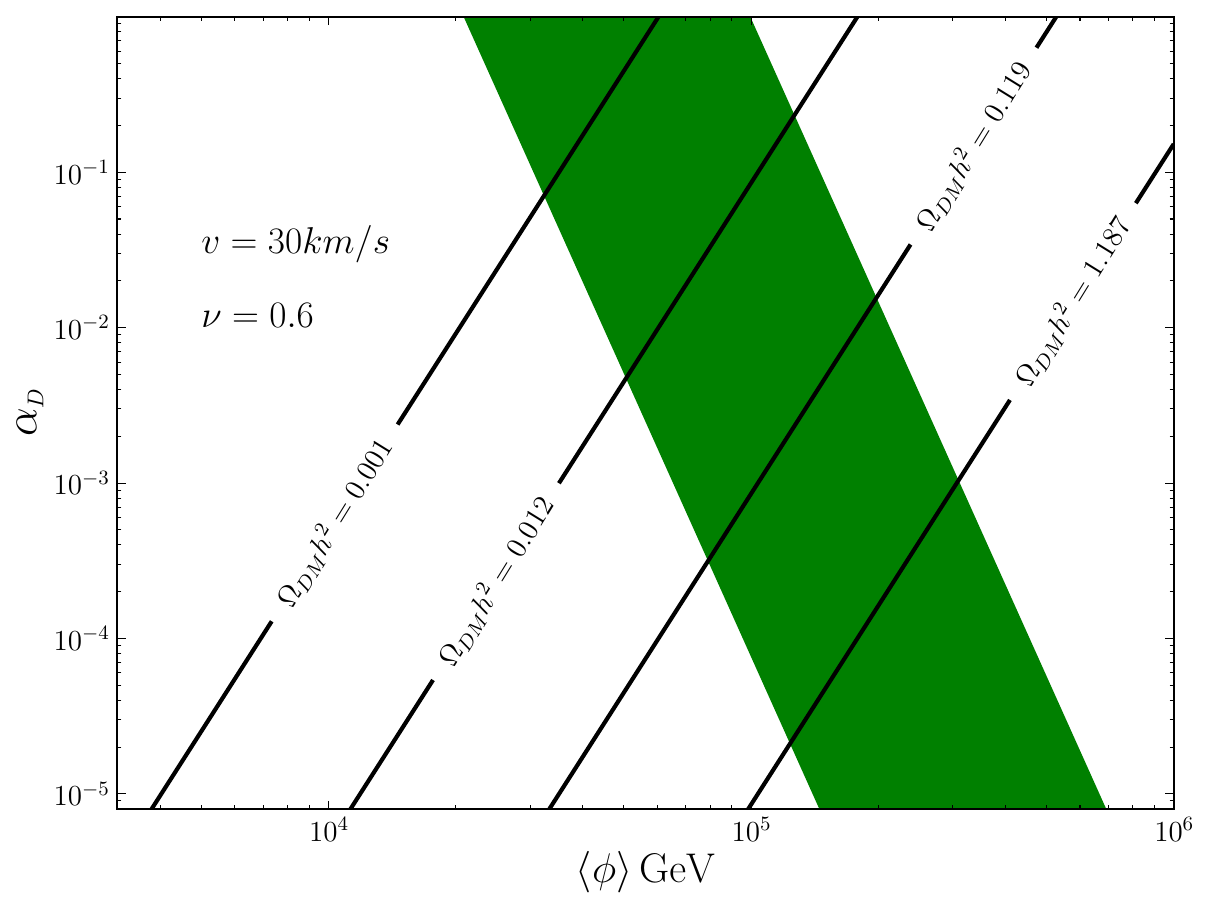}
\caption{Monopole Dark Matter as in Figure \ref{fig:monopole_limit_05} with the critical exponent value $\nu=0.6$.}
\label{fig:monopole_limit_06}
\end{center}
\end{figure}

To solve the too-big-to-fail problem one needs a cross section of the order of $\sigma_{\rm T}/M_{\rm DM}\sim 0.1-10$ cm$^2$/g ~\cite{Zavala:2012us,Tulin:2013teo,Buckley:2014hja} at the velocity scale of drwarf galaxies($v \sim 10-30$ km/s). Comparing this to the limits from larger scale structures one finds that there might be a small region of parameter space left for a theory with velocity independent cross section of around $\sigma_{\rm T}/M_{\rm DM}\sim 0.6$ cm$^2$/g ~\cite{Zavala:2012us,Tulin:2013teo}

In this paper we consider a velocity dependent cross section where if the cross section is around 1 cm$^2$/g at velocities of $v \sim 10-30$ km/s it will be much smallerat the velocities relevant for the shapes of galaxies or the bullet cluster. Therefore there is no contradiction between the cross sections needed to solve the too-big-too-fail problem and the constraints from the elipticity of galaxies.

In Figure \ref{fig:vdm_limit} we show the region in the parameter space of our model, where the transfer cross-section for Vector Dark Matter
is in the desired region $\sigma_{\rm T}/M_{\rm DM} = 0.1-10$ cm$^2$/g at $v -30$ km/s
which can help solving these problems with dwarf galaxies. 
This is superimposed with the contours of the relic density for Vector Dark Matter in our model.  
In Figure \ref{fig:vdm_limit2} we overlay this with the contours of $\sigma_{\rm T}/m_{DM} =1$ at other velocities. 
It readily follows from these considerations that the upper bound constraint from the Milky Way and from Bullet cluster at $v\sim $ 200 to km/s are satisfied
by the self-interacting VDM in the regime where the relic density is in agreement with observations and the dwarf-galaxy-scale problems are addressed.

Monopoles dark matter self-interactions are obtained by replacing the electric with the magnetic Coulomb law,
 $\alpha_{\sst D} \rightarrow \alpha_{\sst mD}=1/\alpha_{\sst D}$, which gives the limits seen in 
 Figs.~\ref{fig:monopole_limit_05} and \ref{fig:monopole_limit_06} 
 for Monopole Dark Matter produced in a model with a second order phase transition. 

\section{Conclusions}

 We have shown that both dark monopoles and dark vector bosons can contribute to and accommodate the observed relic density and that the
 dark long range forces acting upon them correctly satisfy the observational constraints on $N_{\rm eff}$ and on the transfer
 cross-section at large velocities. At the same time, the self-interacting Vector and Monopole DM intrinsic to our model 
 produce the right size of transfer cross-sections relevant for addressing problems with dwarf galaxies.
It would be interesting to use these general features and ingredients in simulations for formation and evolution of structure from the dwarf galaxies scale to
the large scale.

\bigskip

\section*{Acknowledgments}
We would like to thank Chris McCabe for useful discussions and comments.
This material is based upon work supported  by 
STFC through the IPPP grant ST/G000905/1. VVK acknowledges the support of the Wolfson Foundation and Royal Society
through a Wolfson Research Merit Award.  GR acknowledges the receipt of a Durham Doctoral Studentship.

\clearpage

\startappendix
\Appendix{Coleman-Weinberg with an adjoint scalar}

\medskip

\noindent Consider a CSI SU(2)$_{\sst D}\times$SM model with a scalar adjoint field in the hidden sector. The classically massless
SU(2) theory with an adjoint scalar \eqref{eq:LD} was in fact one of the examples considered in the original paper of
Coleman and Weinberg \cite{CW}.
 In a gauge where $\phi_{1,2}=0,\phi_3=\phi$ they find a contribution from the gauge bosons to the effective potential of the form,
\begin{equation}
V_{W'}=\frac{3g_{\sst D}^4}{32\pi^2}\phi^4 \left(\log  \frac{\phi^2}{\bra\phi\ket^2}-\frac{25}{6}\right)
\label{eq:wprime}
\end{equation}
This is twice the result of the Abelian U(1) case as there now two massive vector bosons, $W'_{\pm}$.
This is also to be compared with the case of SU(2)$_{\sst D}$ with a {\it  fundamental} scalar considered in \cite{Khoze:2014xha},
where all three gauge bosons got a mass. Combining the 1-loop expression \eqref{eq:wprime} with tree level potential we get,
\begin{equation}
V= \frac{\lambda_\phi}{4}\phi^4+\frac{3g_{\sst D}^4}{32\pi^2}\phi^4 \left(\log  \frac{\phi^2}{\bra\phi\ket^2}-\frac{25}{6}\right)\,,
\end{equation}
which has a non-trivial minimum with a vev for $\phi$ when 
\begin{equation}
\lambda_\phi(\langle \phi \rangle)\,=\,\frac{11}{8\pi^2}\, g_{\sst D}^4(\langle \phi \rangle)\,.
\end{equation}
With the adjoint scalar having acquired a vev,  the SU(2)$_{\sst D}$ gauge group is broken to U(1))$_{\sst D}$ and 
we end up with two massive gauge bosons $W_{\pm}'$ a massless gauge boson$\gamma'$, and one massive scalar field $\phi=\phi_3$
neutral under U(1))$_{\sst D}$ .
The masses are given by:
\begin{equation}
M_{W'}=g_{\sst D} \bra\phi\ket\,, \qquad m^2_\phi=\frac{3g_{\sst D}^4\bra\phi\ket^2}{4\pi^2}\,.
\end{equation}

\medskip
We now include the effects of the portal coupling between $\phi$ and the Higgs. The scalar potential is now,
\begin{equation}
V=\lambda_H |H|^4 + \lambda_\phi Tr(\Phi \Phi^\dagger)^2-\lambda_{\rm P}|H|^2Tr(\Phi \Phi^\dagger)\,.
\end{equation}
When $\phi$ develops a vev the portal term acts as a negative mass term for the Higgs and triggers electroweak symmetry breaking. 
Some of the relations above also get contributions from $\lambda_{\rm P}$
\begin{eqnarray}
\lambda_\phi &=& \frac{11}{16\pi^2} \,g_{\sst D}^4 
+\lambda_{\rm P}\frac{v^2}{2\langle\phi\rangle^2}
\qquad {\rm at} \quad \mu=\langle \phi\rangle\,,
\label{eq:cwmsbar-P}
\\
m_\phi^2 &=&
\frac{3g_{\sst D}^4\bra\phi\ket^2}{4\pi^2}
+\lambda_{\rm P} v^2\,.
\label{eq:mphiZ-P}
\end{eqnarray}
The Higgs portal interaction is also responsible for the mixing between the SM Higgs $h$ and the $\phi$ scalar of the Dark sector,
\begin{equation}
\label{Mmixing}
  M^{2}=\left(
\begin{array}{cc}
2\lambda_H\, v^2   &  - \sqrt{2\lambda_{\rm P} \lambda{H}} v^2
\\
% & \\
 - \sqrt{2\lambda_{\rm P} \lambda{H}} v^2  &  m^{2}_{\phi} 
\end{array}\right)\,.
\end{equation}

%\clearpage

\bibliographystyle{h-physrev5}

\end{document}